\documentclass{caosp309}

\usepackage{graphicx}

\usepackage{natbib}
\usepackage{amssymb}

\renewcommand{\cm}[0]{\multicolumn{1}{c}{\checkmark}}
\newcommand{\mcm}[0]{\multicolumn{1}{c}{(\checkmark)}}
\newcommand{\rev}[1]{}

\articleNo{301}
\pubyear{2020}
\volume{50}
\volnumber{3}
\firstpage{1}
\received{May 1, 2020}
\accepted{July 28, 2020}

\begin{document}

\hauthor{A.~Pr\v sa}

\title{The present and the future of modeling eclipsing binary systems}

\author{
    A.\,Pr\v sa\inst{1}\orcid{0000-0002-9138-9028}
}

\institute{
    Villanova University, Department of Astrophysics and Planetary Science, 800 Lancaster Ave, Villanova PA 19085, \email{aprsa@villanova.edu}
}

\date{December 15, 2024}

\maketitle

\begin{abstract}
In September 2024, eclipsing binary star practitioners gathered in Litomy\v sl, Czech Republic, the birth town of Zde\~ nek Kopal, one of the most celebrated pioneers of our field, to discuss the latest developments and state-of-the-art. I was invited to present my own biased view of the present and the future of modeling eclipsing binary stars. In this contribution I attempt to make a clear distinction between approaches that are suited to individual objects and approaches that aim to deliver bulk results for large datasets. I stress that our motivation should be different: individual system analysis is warranted whenever there is potential to propose or improve our understanding of the underlying physics, while bulk analysis should be used to probe stellar formation and evolution channels. I briefly discuss two examples of tools to achieve the goals: PHOEBE for individual system analysis, and PHOEBAI for bulk analysis.

\keywords{binary stars -- eclipsing binaries -- fundamental parameters -- artificial intelligence}
\end{abstract}

\section{Introduction} \label{sec_introduction}

\begin{figure}
    \centering
    \includegraphics[width=0.425\textwidth]{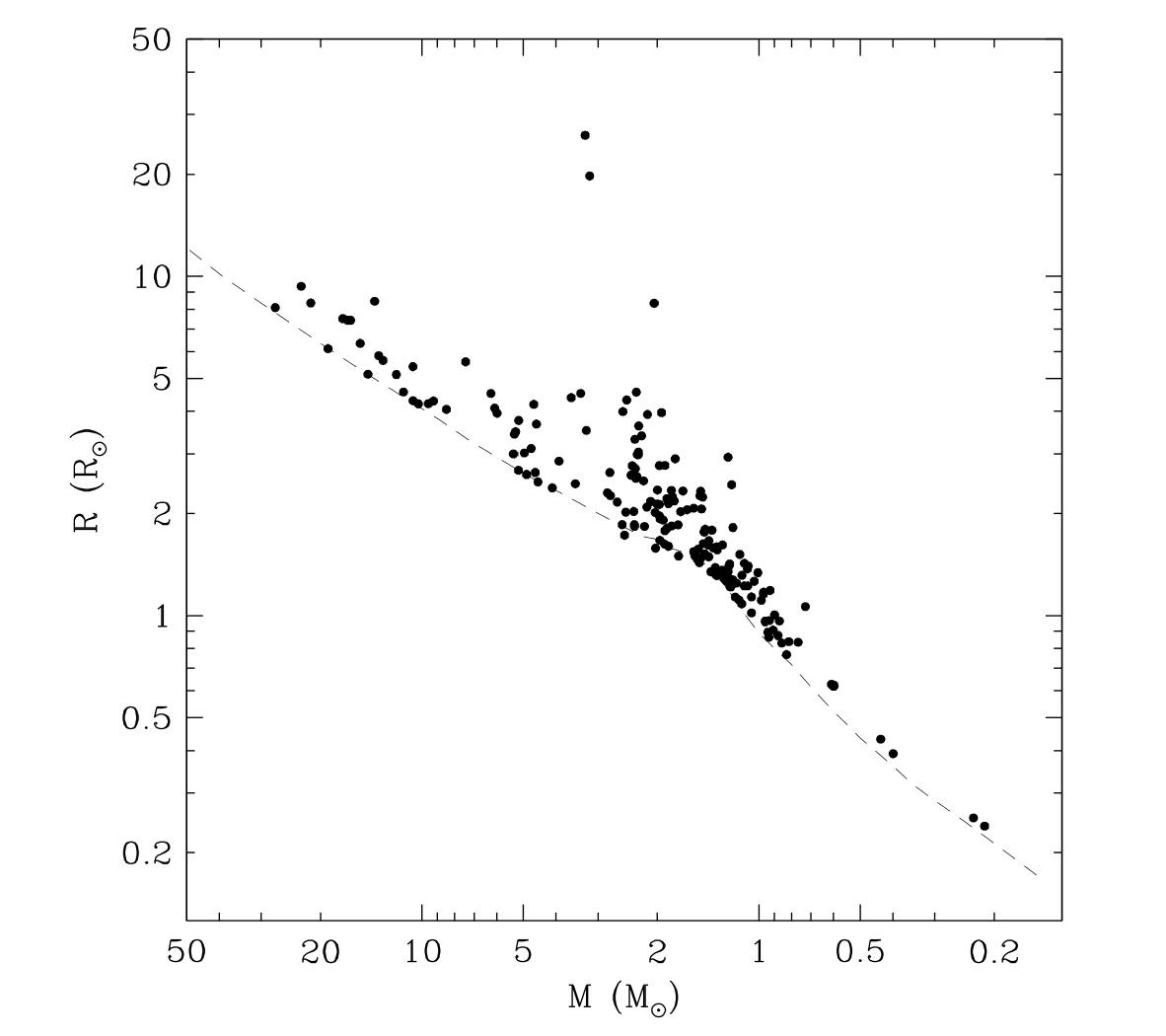}
    \includegraphics[width=0.565\textwidth]{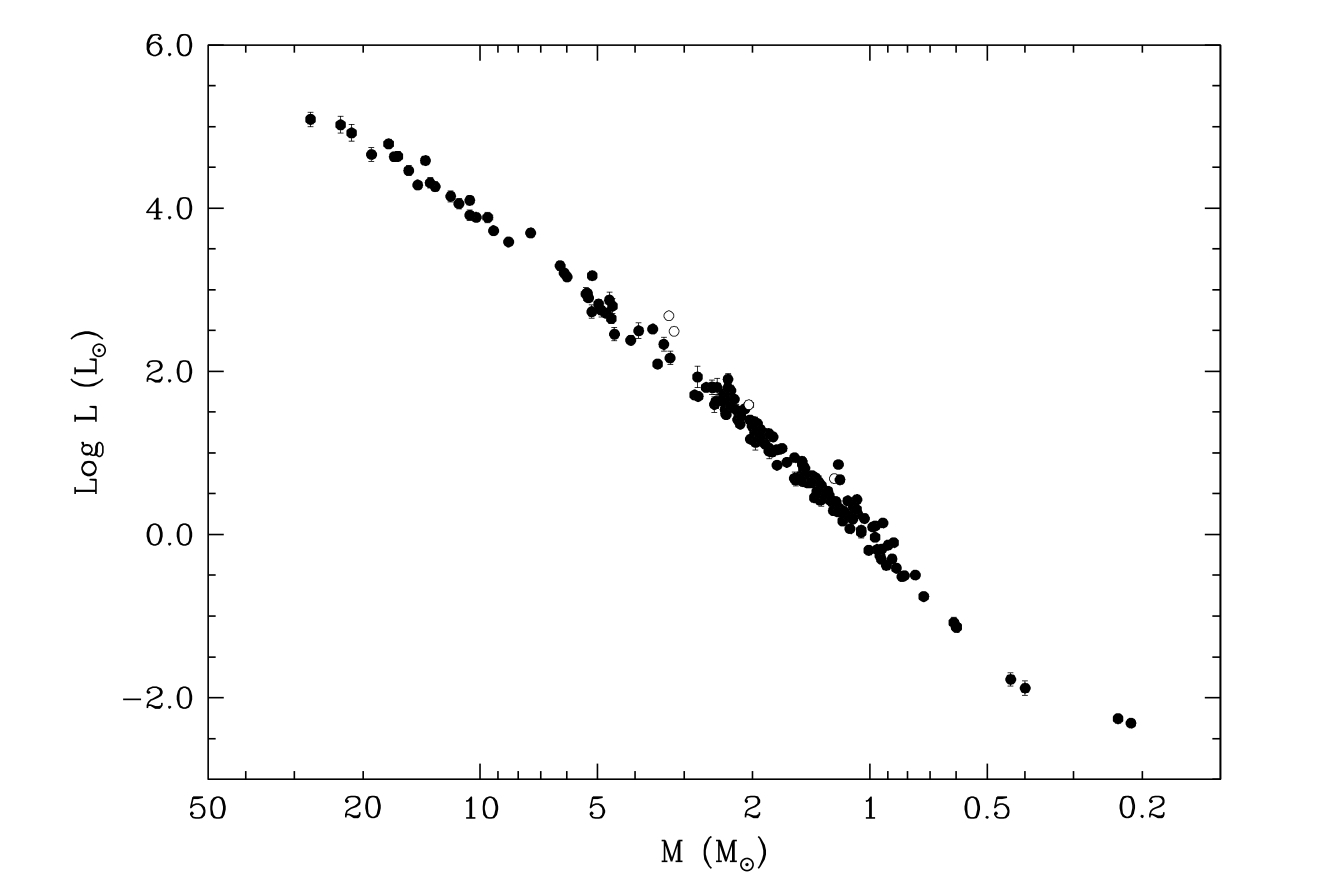}
    \caption{Radius-mass and luminosity-mass relationships for a sample of 94 EBs \citep{torres2010}. The dashed line is the theoretical zero-age main sequence. Uncertainties in $R$, $L$ and $M$ are smaller than symbol sizes. Depicted with open symbols are stars classified as giants.}
    \label{fig:abspars}
\end{figure}

Eclipsing binary stars (EBs) have long been recognized as \emph{the} calibrators of stellar astrophysics. Most of what we know about the fundamental stellar properties \rev{(masses, radii, temperatures and luminosities)} stems from EB studies \citep{torres2010}. Their favorable orbital alignment with the line of sight, and consequent eclipses, make them ideal astrophysical laboratories: a simple geometry coupled with well-understood dynamical laws allow us to obtain fundamental parameters without a-priori assumptions. Fig.~\ref{fig:abspars} showcases the power of EBs: the masses, radii and luminosities of stars inferred from EB observations constitute the most accurate set of constraints for the theories of stellar formation and evolution. The uncertainties are generally below 2-3\%, making the error-bars in Fig.~\ref{fig:abspars} smaller than the symbols.

Binary systems do not discriminate on the spectral type or luminosity class; be it main sequence stars, red giants, or compact objects, be it flaring M-dwarfs or pulsating variables, be it tight mass-transferring systems or exoplanets, all these components are found in binary stars. Once we find them in an eclipsing system, the path to fundamental parameters is straight-forward. For obvious reasons, colleagues who study these objects rely on masses and radii provided from EB analyses, which earns EB practitioners an occasional bottle of wine for their efforts. Further augmenting the importance of binaries is the observation that around 20\% of binaries have tertiary components \citep{orosz2015}, including circumbinary planets \citep{welsh2015}; when all three components undergo eclipses, the uncertainties in fundamental parameters can reach $\sim$0.1\% \citep{carter2011}. It is thus of no surprise that Henry Norris Russell, in \citeyear{russell1948}, pointed out EBs as ``the Royal Road'' to stellar astrophysics.

\section{The Present} \label{sec:present}

The 21st \rev{century} has been marked by an unprecedented advancement in data quality. NASA's \textsl{Kepler} mission, for example, routinely achieved photometric precision of the order of few tens of ppm \citep{borucki2010}. \textsl{Kepler} observed $\sim$150,000 stars in the 105 square degree patch of the sky for 4 years with a 30-min cadence (and $\sim$1,000 stars with 2-min cadence) with 92\% temporal completeness. In effect, for the first time we obtained a near-uninterrupted, ultra-precise set of observations for a few thousand EBs as well \citep{kirk2016}. \textsl{Kepler}'s successor, \textsl{TESS} \citep{ricker2015}, continued the census with thousands of targeted EBs \citep{prsa2022} and $\sim$150,000 EBs in full frame images across the sky. While the majority of \textsl{TESS} EBs are observed continuously for about a month at a time, the dataset still represents a glimpse into EB complexities that remained hidden from ground-based observatories before. The latest space survey, Gaia \citep{gaia2021}, provides a stereoscopic vision of the Galaxy with simultaneous photometric, radial velocity, and astrometric observations, with impressive first EB results for 400,000 systems \citep{mowlavi2023}.

This explosion in photometric precision and temporal coverage put EB models to the test. At this level of precision, the models could no longer reliably reproduce observations, in part because of the missing physics, and in part because of the lack of the required numerical precision. PHOEBE \citep{prsa2005} is one of the many codes that exhibited these inadequacies; this prompted a code rewrite in \citeyear{prsa2016} to remediate the situation.

Irrespective of the choice for the utilized model, the new decade brought on new requirements for modeling. EB practitioners have long realized that -- depending on the types of observables available -- the parameter space is non-linear and severely degenerate. Table \ref{tab:params} provides rule-of-thumb information content for available types of observables: it informs us what parameters we can infer from what types of observations. Thus, in general, to infer fundamental parameters (masses, radii, temperatures and luminosities), we need multi-band photometric and double-lined spectroscopic data.

\begin{table}[t!]
\centering
\footnotesize
\begin{tabular}{l|p{1.3cm}p{1.3cm}p{1.3cm}p{1.3cm}p{1.3cm}p{1.3cm}}
Parameter & Cal.~LC & \mbox{Cal.~LC}+ distance & \mbox{$2$+ Cal.} LCs & \mbox{$2$+ Cal.} LCs+dist. & \mbox{$2$+ Cal.} LCs+SB1 & \mbox{$2$+ Cal.} LCs+SB2 \\
\hline
$t_0$                         & \cm  & \cm  & \cm  & \cm  & \cm  & \cm  \\
$P_0$                           & \cm  & \cm  & \cm  & \cm  & \cm  & \cm  \\
$i$                           & \cm  & \cm  & \cm  & \cm  & \cm  & \cm  \\
$e \cos \omega$        & \cm  & \cm  & \cm  & \cm  & \cm  & \cm  \\
$e \sin \omega$            & \cm  & \cm  & \cm  & \cm  & \cm  & \cm  \\
$r_1+r_2$              & \cm  & \cm  & \cm  & \cm  & \cm  & \cm  \\
$T_2/T_1$                    & \cm  & \cm  & \cm  & \cm  & \cm  & \cm  \\
$r_2/r_1$              & \mcm & \mcm & \mcm & \mcm & \mcm & \cm  \\
$q = M_2/M_1$                       & \mcm & \mcm & \mcm & \mcm & \mcm & \cm  \\
$l_3$                              & \mcm & \mcm & \mcm & \mcm & \mcm & \mcm \\
$\dot P$, $\dot \omega$, \dots & \mcm & \mcm & \mcm & \mcm & \mcm & \mcm \\
$L_1 + L_2$           &      & \cm  &      & \cm  &      & \cm  \\
$L_1$, $L_2$             &      & \cm  &      & \cm  &      & \cm  \\
$T_1$, $T_2$             &      &      & \cm  & \cm  & \cm  & \cm  \\
$v_\gamma$                &      &      &      &      & \cm  & \cm  \\
$M_1$, $M_2$                   &      &      &      &      &      & \cm  \\
$R_1$, $R_2$                    &      &      &      & \cm  &      & \cm  \\
$a$                            &      &      &      & \cm  &      & \cm  \\
\hline
\end{tabular}
\normalsize
\caption{A table of EB parameters that can be inferred from a specific combination of observables. ``Cal.~LC'' stands for flux-calibrated light curve, ``SB1'' stands for a single-lined spectroscopic binary, and ``SB2'' for the double-lined spectroscopic binary. The (\checkmark) symbol corresponds to quantities that can be inferred from some but not all light curves. Adapted from \citet{wrona2024}.}
\label{tab:params}
\end{table}

Further, gone are the days where ``solving'' an EB means running an optimizer that minimizes the sum of squares of the residuals between the data and the model, providing (unrealistically small) formal errors and publishing the results based on poorly sampled data with large scatter. Instead, the typical solution process involves several steps \citep{conroy2020}:

\begin{description}
    \item[Solution estimation:] in order to ascertain convergence for the optimizers, the starting point in the parameter space needs to be reasonably close to a solution\footnote{We deliberately stress \emph{a} solution rather than \emph{the} solution because, in most circumstances, there will not be a single, unequivocal model that uniquely explains the data.}. This is a task for estimators: methods that quickly analyze the data and provide a cursory solution. An example is an estimate of $e \cos \omega$ from eclipse separation in phase, or $r_1/a + r_2/a$ from eclipse widths, or $e \sin \omega$ from the eclipse duration ratios.

    \item[Solution optimization:] once close, the solution needs to converge to a minimum as robustly as possible. This is usually achieved by finding the parameters that minimize the sum of squares of the residuals. Optimization is an iterative process that will typically follow the locally steepest slope towards the minimum. Once there, it will estimate formal errors from the covariance matrix. \rev{Examples of} optimizers \rev{include} differential corrections \citep{wilson1971}, Nelder and Mead's simplex method \citep{kallrath1987}, and Powell's direction set method \citep{prsa2007}.

    \item[Solution sampling:] the minimum reached must not be construed as the final solution: the topology of the parameter space near that minimum needs to be properly explored. That is a task for samplers: methods that traverse the vicinity of the parameter space and evaluate the probabilities in order to infer realistic parameter values. This is a computationally expensive and arduous process as parameter space needs to be traversed heuristically. Typical samplers include Markov Chain Monte Carlo methods \citep{emcee2019}, differential evolution \citep{storn1997}, etc. Inference is usually done within the Bayesian formalism, and the results are typically presented in a corner plot, with individual parameter posterior probability density functions and two-dimensional parameter cross-correlations \citep{conroy2020}.
\end{description}

These components need to be utilized in unison in order to derive a compelling, credible solution for EB parameters given the data. The time cost of a typical modeling process is weeks to months, depending on the complexity of the parameter space, types of observations, and data quality and quantity. \rev{To find a minimum (for example, by using Nelder and Mead's method) and adequately sample the parameter space (for example, by using MCMC with 100 walkers and 10,000 iterations) for a detached binary star, the typical number of forward model evaluations is of the order of $10^6$.}

\section{The Future} \label{sec:future}

The 21st \rev{century} has also been marked by an unprecedented advancement in data quantity. Table \ref{tab:ebno} provides a rough census of observed or expected numbers of EBs in recent, currently ongoing, and imminent surveys.

\begin{table}[t!]
    \centering
    \begin{tabular}{llll}
        \hline
        Survey & Data type & Years of operation & EB yield \\
        \hline
        K2       & photometry                  & 2014-18   & $\sim$1,000       \\
        Kepler   & photometry                  & 2009-13   & $\sim$3,000       \\
        NGTS     & photometry                  & 2015--    & $\sim$5,000       \\
        OGLE-II  & photometry                  & 1997-2000 & $\sim$6,000       \\
        Plato    & photometry                  & 2026?     & $\sim$8,000       \\
        ASAS     & photometry                  & 1997-2010 & $\sim$10,000      \\
        Galah    & spectroscopy                & 2013--    & $\sim$15,000      \\
        OGLE-III & photometry                  & 2001-09   & $\sim$40,000      \\
        ASAS-SN  & photometry                  & 2014--    & $\sim$40,000      \\
        TESS     & photometry                  & 2018--    & $\sim$150,000     \\
        OGLE-IV  & photometry                  & 2010--    & $\sim$500,000     \\
        ZTF      & multi-band photometry       & 2018--    & $\sim$500,000     \\
        Gaia     & photometry, astrometry, RVs & 2013--    & $\sim$1.8 million \\
        CSST     & photometry, spectroscopy    & 2026?     & $\sim$2 million   \\
        LSST     & multi-band photometry       & 2025--    & $\sim$10 million  \\
        \hline
    \end{tabular}
    \caption{A census of recent, ongoing, and imminent surveys, along with their EB yield.}
    \label{tab:ebno}
\end{table}

It seems clear that, over the next ten years or so, we will have well over ten million EBs in our hands. In the previous section we have established that, typically, we need $\sim$weeks of computer time to robustly analyze EB observables. Thus, unless we are willing to accept that EB modeling remains a boutique\footnote{The task of solving 10 million EBs in the next 100 years would require $\sim$4000 astronomers and $\sim$8 million computer cores, along with a moratorium on further observations.} operation, we need faster approaches. 

One such approach utilizes artificial intelligence (AI). We have been inundated with AI in our everyday lives, and for good reason. It powers technologies like virtual assistants, personalized recommendations, and smart home devices, making daily tasks more efficient and tailored to individual needs. In healthcare, AI enables early diagnosis through advanced imaging and predictive analytics \citep{secinaro2021}, while in education, it supports personalized learning experiences \citep{chen2020}. AI also enhances industries like transportation with autonomous vehicles and navigation systems, and it revolutionizes customer service through chatbots and automated support \citep{mohamad2021}.

AI has been proposed for EBs as well, for automated classification (see, for example, \citealt{cokina2021}) and for bulk analysis (e.g.,~\citealt{prsa2008}). The appeal is undeniable: if AI could ``look at'' observables, recognize them as EBs, and deduce principal parameters in a fraction of a second, it would allow us to rapidly analyze large datasets. The question, then, is how accurately can this be done? \citet{prsa2008} trained a backpropagating neural network on synthetic light curves generated by PHOEBE \citep{prsa2005,prsa2016} where the network was shown $\sim$32,000 light curves with 5 corresponding parameters: temperature ratio ($T_2/T_1$), sum of fractional radii ($R_1/a + R_2/a$), radial eccentricity ($e \sin \omega$), tangential eccentricity ($e \cos \omega$), and inclination ($\sin i$). They demonstrated that the trained network could attain the accuracy better than 10\% for 90\% of the sample. This allowed the authors to deploy the network on 2500+ EBs from OGLE and CALEB databases.

The principal problem with this type of approach is that, in the absence of post-festum tests, we do not know \emph{which} 10\% are not correctly characterized. Given that the neural network maps input (light curves) to output (parameters) opaquely, not only do we lack the ability to evaluate per-light curve accuracy, we also cannot attribute meaningful uncertainties to any obtained parameters. The remedy is to run post-festum tests, but these can be both computationally and temporally expensive, which defeats the initial purpose at least to some extent.

\begin{figure}
    \centering
    \includegraphics[width=\textwidth]{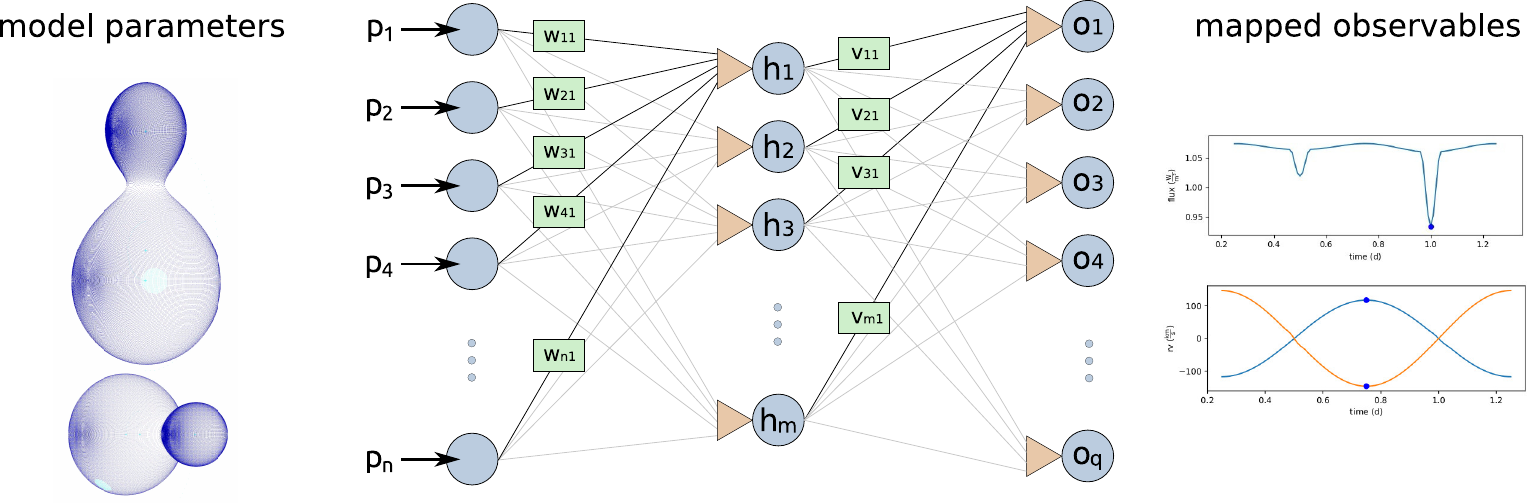} \\
    \caption{A schematic representation of the feed-forward neural network that acts as an emulator to the physical engine. Parameters $p_i$ are mapped through connection weights $w_{ij}$ across hidden layers $h_j$ to the output units $o_k$, representing synthetic observables.}
    \label{fig:nn}
\end{figure}

There exists an alternative, though: we can exchange the neural network's inputs and outputs. Instead of providing light curves as input, we provide parameters; instead of collecting parameters as output, we collect light curves. The network becomes an \emph{emulator} of the physical engine. Assuming that the network can be trained adequately to encompass the non-linear properties of our complex, multivariate parameter space, each computation that takes minutes to complete using the physical model can take milliseconds using a trained network. Thus, the AI-powered engine becomes a drop-in replacement for the physical model. In our particular case, PHOEBAI replaces PHOEBE (cf.~Fig.~\ref{fig:nn}).

Let us first estimate the potential speed-up and establish a good motive for this effort. In Section \ref{sec:present} we estimate that, for the completion of the estimation, optimization, and sampling stages, we require $\sim 10^6$ forward model computations. If each model takes $\sim$2 minutes to compute \rev{(typical PHOEBE runtime per sector of \textsl{TESS} short cadence data for an eccentric, time-variable EB)}, we are talking about 3.8 \emph{years} of compute time on a single processor core. To handle this volume of computations in a reasonable time (say, 2 weeks), we need 100 cores -- i.e., a computer cluster. For the neural network, on the other hand, the model can be ``computed'' within a millisecond. In that case, the net compute time becomes $\sim$1.5 hours \emph{on a single processor}. \rev{The training time for the neural network, while appreciable, is a one-time effort and does not impact any subsequent runtime costs.} Thus, AI holds promise of a $\sim10^6$-fold speed-up.

There are certainly limitations to such an emulator, even if the underlying network is perfectly trained. 
\begin{itemize}
    \item The inputs and the outputs are fixed, determined by the network topology. If the network it trained with, say, 10 input parameters and it outputs, say, 500 fluxes in equidistant phases, we are restricted to using the same 10 parameters and the emulator will yield fluxes in those exact 500 phase points. The observations will thus need to be interpolated to the corresponding phase array for comparison with the model, and any other parameters present in the physical model will not play a role in the emulator.

    \item The network emulates phased light curves, so there can be no temporal variation in the data. Any variability that is not captured by the parameters used for training the network will inevitably skew the results to the extent of the degeneracy between included and omitted parameters. Any trends in the data, or more complex noise models, need to be accounted for either before (i.e., by detrending) or concurrently with sampling (i.e., by modifying the maximized probability function).
    
    \item Neural networks cannot be used on input that deviates from the training set: while they are well suited for interpolation, they are notoriously bad at extrapolation \citep{freeman1991}. Thus, if the training set does not cover the parameter space adequately, sampling will not be done correctly. It also means that, if the \emph{density} of the covered parameter space is not representative of actual distributions, the results may be biased or suffer from undersampling systematics.

    \item The parameter space is inherently non-linear and highly degenerate. The choice of input parameters is thus crucial for emulator performance and result fidelity. Should any of the principal parameters be omitted, or if the chosen parameters are not orthogonal, the emulator will be misspecified and the results will be biased.
\end{itemize}

These limitations can be controlled with a careful optimization of the neural network topology and the construction of the training set. Once optimized and trained, the emulator is ready to be used as a drop-in replacement of the physical model. That includes the optimization stage and the sampling stage: given the data, the emulator is able to minimize the cost function and to sample the probability density. It is able to do so in a $\sim$millionth of the physical model's time cost.

As proof of concept, we trained an emulator on 6 photometric parameters: the effective temperature ratio $T_2/T_1$, eccentricity $e$, argument of periastron $\omega$, orbital inclination $i$, the sum of fractional radii $r_1+r_2$, and their ratio $r_2/r_1$. For efficiency, we replaced $e$ and $\omega$ with $e \sin \omega$ and $e \cos \omega$, and $i$ with $\cos i$. The parameters were sampled from distributions that cover a wide enough range to encompass the case study light curves. The network was trained on $\sim$600k light curves synthesized using PHOEBE. As outputs, the network used phase-folded light curves sampled in 500 points. We denote this emulator PHOEBAI: PHOEBE via AI \citep{wrona2024}.

The topology of the network was determined by the \texttt{RandomizedSearchCV} method \citep{bergstra2012} from the \texttt{sklearn} library \citep{sklearn2011}. The method evaluated thousands of structures with varying numbers of hidden layers, nodes, and activation functions. The chosen shape was the simplest among the best performing in terms of minimizing the sum of residual squares between predicted and actual sets of light curves. It comprises four hidden layers with 512, 512, 512, and 1024 nodes, with the sigmoid activation function.

Application of this network was tested on a subset of EBs from the \textsl{TESS} EB catalog \citep{prsa2022} that matched our training set distributions. The optimizer utilized differential evolution \citep{storn1997} from the \texttt{scipy.optimize} library \citep{scipy2020}. The final results from this optimization were subsequently used as starting values for MCMC sampling using the emcee Python library \citep{emcee2019}.

\begin{figure}[p!]
    \centering
    \includegraphics[width=\textwidth]{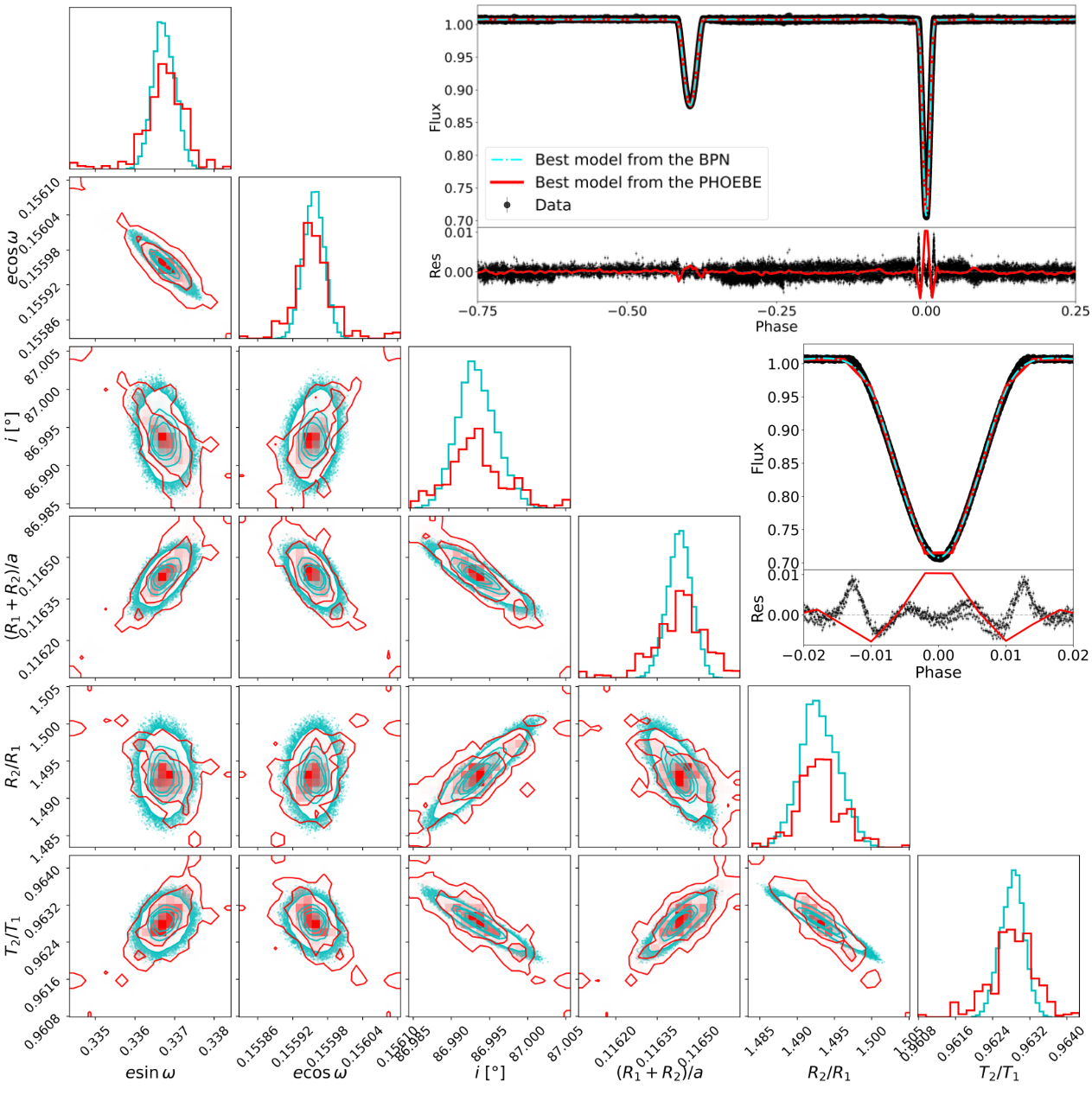}
    \caption{Sampling results for TIC 279097963, a detached EB observed by \textsl{TESS} in Sector 4. The top right panel depicts a phase-folded light curve with the best-fit PHOEBE model (red) and the best-fit PHOEBAI model (cyan), along with the residuals, plotted over data. The panel below zooms in on the primary eclipse, demonstrating that the residuals are due to the coarse phase sampling. The corner plot compares parameter posteriors (diagonal) and 2-D correlations (off-diagonal) for the 6 sampled parameters with PHOEBE (red) and PHOEBAI (cyan). PHOEBAI posteriors are better sampled but slightly narrower. Overall, there is a remarkable agreement between the two approaches, where PHOEBAI performed $\mathrm{\sim}250,000$ times faster than the PHOEBE sampler.}
    \label{fig:comp}
\end{figure}

During the sampling process, we employed 80 walkers and 4000 iterations to obtain posterior distributions. Fig.~\ref{fig:comp} compares PHOEBAI and PHOEBE performance on a detached EB, TIC 279097693. This is a typical (rather than cherry-picked) result from the test sample: it shows that PHOEBAI is able to accurately capture both the posteriors and parameter correlations.

\section{Conclusions}

As we continue to be inundated with high-quality data, we need to distinguish between ``boutique'' operation and bulk processing. When our understanding of the physical processes can be advanced by pointed investigations of individual objects, the well-established methods and codes remain indispensable. On the other hand, to understand large sample properties and parameter distributions that probe the outcomes of stellar formation and evolution principles, we need to rely on novel methods. One such method, presented in this paper, employs feed-forward neural networks. It is important to realize that it is not the only way or even the preferred way to model large swaths of data, but -- as demonstrated here -- PHOEBAI certainly shows promise in delivering robust results on par with the physical engines. Above all, it remains vital to understand the limitations and ranges of application for both the physics-based models and the AI-based emulators and to use them appropriately. 

\acknowledgements

The author would like to acknowledge NASA award 23-ADAP23-0068 that funded in part the work presented in this paper.

\bibliography{prsa}

\end{document}